\documentclass[aps,prl,twocolumn,10pt,showpacs,superscriptaddress,groupedaddress,longbibliography]{revtex4-1} 
\pdfoutput=1

\usepackage{graphicx}        
\usepackage{dcolumn}        
\usepackage{bm}             
\usepackage{amssymb}
\usepackage{amsmath}
\usepackage{epstopdf} 
\usepackage{color}
\definecolor{korr_26Apr}{rgb}{0,0,0} 
\definecolor{red}{rgb}{1,0,0}

\def \d{\mathrm{d}}

\begin{document}

\widetext

\title{Unification of Aeolian and Fluvial Sediment Transport Rate from Granular Physics}
\author{Thomas P\"ahtz$^{1,2}$}
\email{0012136@zju.edu.cn}
\author{Orencio Dur\'an$^3$}
\affiliation{1.~Institute of Port, Coastal and Offshore Engineering, Ocean College, Zhejiang University, 310058 Hangzhou, China \\
2.~State Key Laboratory of Satellite Ocean Environment Dynamics, Second Institute of Oceanography, 310012 Hangzhou, China \\
3.~Department of Ocean Engineering, Texas A\&M University, College Station, Texas 77843-3136, USA}

\begin{abstract}
One of the physically least understood characteristics of geophysical transport of sediments along sediment surfaces is the well known experimental observation that the sediment transport rate $Q$ is linearly dependent on the fluid shear stress $\tau$ applied onto the surface in air, but is nonlinearly dependent on $\tau$ in water. Using transport simulations for a wide range of driving conditions, we show that the scaling depends on the manner in which the kinetic fluctuation energy of transported particles is dissipated: via predominantly fluid drag and quasistatic contacts (linear) versus fluid drag and quasistatic and collisional contacts (nonlinear). We use this finding to derive a scaling law (asymptotically $Q\sim\tau^2$) in simultaneous agreement with measurements in water and air streams.
\end{abstract}

\maketitle

Turbulent shearing flows of Newtonian fluid along planetary surfaces composed of loose sediment are an important driver of sediment transport and erosion on Earth and other planets~\cite{Pahtzetal20a,Garcia08,Ancey20,Duranetal11,Koketal12,Bagnold41}. In particular, if the transported sediment is frequently deposited on the sediment bed underneath (i.e., if it is not suspended by the fluid turbulence), then the interplay between erosion, deposition, bed topography, and flow gives rise to a rich variety of bedforms, such as desert dunes and subaqueous ripples~\cite{Bagnold41,Bourkeetal10,Charruetal13}. Predicting the evolution of fluid-sheared planetary surfaces thus requires a profound understanding of nonsuspended sediment transport~\cite{ClaudinAndreotti06,Pahtzetal13,Pahtzetal14,Pahtzetal15b}, especially of the dependency of the transport rate $Q$ on environmental parameters, such as the fluid shear stress $\tau$ applied onto the bed. Measurements have revealed that $Q$ scales approximately linearly with $\tau$ in aeolian (i.e., air-driven) transport~\cite{Creysselsetal09,Hoetal11,MartinKok17}, but scales nonlinearly with $\tau$ in fluvial (i.e., liquid-driven) transport~\cite{MeyerPeterMuller48,SmartJaeggi83,Gao08,CapartFraccarollo11}. However, the physical origin of this difference remains controversial~\cite{Berzietal16,PahtzDuran18a,PahtzDuran18b} and a general scaling law for $Q$ elusive.

Here, using discrete element method-based sediment transport simulations, we show that the linear-to-nonlinear transition in the scaling of $Q$ with $\tau$ is caused by a regime shift in the manner in which kinetic fluctuation energy of transported particles is dissipated. Via parametrizing this shift, we derive a general scaling law, valid for continuous (not intermittent) turbulent transport of nearly monodisperse sediment, in simultaneous agreement with measurements in water and air streams.

We use the numerical model of Ref.~\cite{Duranetal12}, which couples a discrete element method for the particle motion under gravity, buoyancy, and fluid drag with a continuum Reynolds-averaged description of hydrodynamics. Spherical particles ($\sim 10^4$) with mild polydispersity are confined in a quasi-two-dimensional domain of length $\sim 10^3d$ (where $d$ is the mean particle diameter), with periodic boundary conditions in the flow direction, and interact via normal repulsion (restitution coefficient $e=0.9$) and tangential friction (contact friction coefficient $\mu_c=0.5$). The bottom-most particle layer is glued on a bottom wall, while the top of the simulation domain is reflective but so high that it is never reached by transported particles. The Reynolds-averaged Navier-Stokes equations are combined with a semiempirical mixing length closure that ensures a smooth hydrodynamic transition from high to low particle concentration at the bed surface and quantitatively reproduces the mean turbulent flow velocity profile in the absence of transport. Simulations with this numerical model are insensitive to $e$, and therefore insensitive to viscous damping, and simultaneously reproduce measurements of the rate and threshold of aeolian and viscous and turbulent fluvial transport (Figs.~1 and 3 of Ref.~\cite{PahtzDuran18a}), height profiles of relevant equilibrium transport properties (Fig.~2 of Ref.~\cite{PahtzDuran18a} and Fig.~6 of Ref.~\cite{Duranetal14a}), and aeolian ripple formation~\cite{Duranetal14b}. Details of the numerical model and its validation are described in the Supplemental Material~\cite{SuppLiquidBed}.

The simulated steady, homogeneous transport conditions are characterized by three dimensionless numbers: the particle-fluid-density ratio $s\equiv\rho_p/\rho_f$, Galileo number $\mathrm{Ga}\equiv\sqrt{s\tilde gd^3}/\nu$ (also known as Yalin parameter~\cite{Yalin77}), and Shields number $\Theta\equiv\tau/(\rho_p\tilde gd)$, where $\tilde g\equiv(1-1/s)g$ is the buoyancy-reduced value of the gravitational constant $g$ and $\nu$ the kinematic fluid viscosity. The density ratio $s$ separates aeolian ($s\gtrsim10$) from fluvial ($s\lesssim10$) conditions, while $\mathrm{Ga}$ controls the dimensionless terminal settling velocity (i.e., $v_s/\sqrt{s\tilde gd}=f(\mathrm{Ga})$)~\cite{PahtzDuran18a}.

We consider a Cartesian coordinate system $(x,y,z)$, with $x$ the horizontal coordinate in the flow direction, $z$ the vertical coordinate in the direction normal to the bed oriented upwards, and $y$ the lateral coordinate. For a given simulated condition, we define the bed surface elevation $z_r$ as the effective elevation of transported particles rebounding with the bed~\cite{PahtzDuran18b} (see the Supplemental Material~\cite{SuppLiquidBed} for the computation of physical quantities from the simulation data). From the masses ($m^i$) and velocities ($\mathbf{v}^i$) of all particles, numbered consecutively (upper index $i$), we then obtain the sediment transport rate $Q\equiv\frac{1}{\Delta}\langle\sum_im^iv^i_x\rangle_T$ (i.e., the average total horizontal particle momentum per unit bed area), transport load $M\equiv\frac{1}{\Delta}\langle\sum_{z^i\geq z_r}m^i\rangle_T$ (i.e., the average total mass of particles transported above $z_r$ per unit bed area), and average horizontal velocity of transported particles $\overline{v_x}\equiv Q/M$, where $\Delta$ is the area of the $(x,y)$-simulation domain and $\langle\cdot\rangle_T$ denotes the time average. These quantities are nondimensionalized via $Q_\ast\equiv Q/(\rho_pd\sqrt{s\tilde gd})$, $M_\ast\equiv M/(\rho_pd)$, and $\overline{v_x}_\ast\equiv\overline{v_x}/\sqrt{s\tilde gd}$.

\begin{figure}[!htb]
 \begin{center}
  \includegraphics[width=1.0\columnwidth]{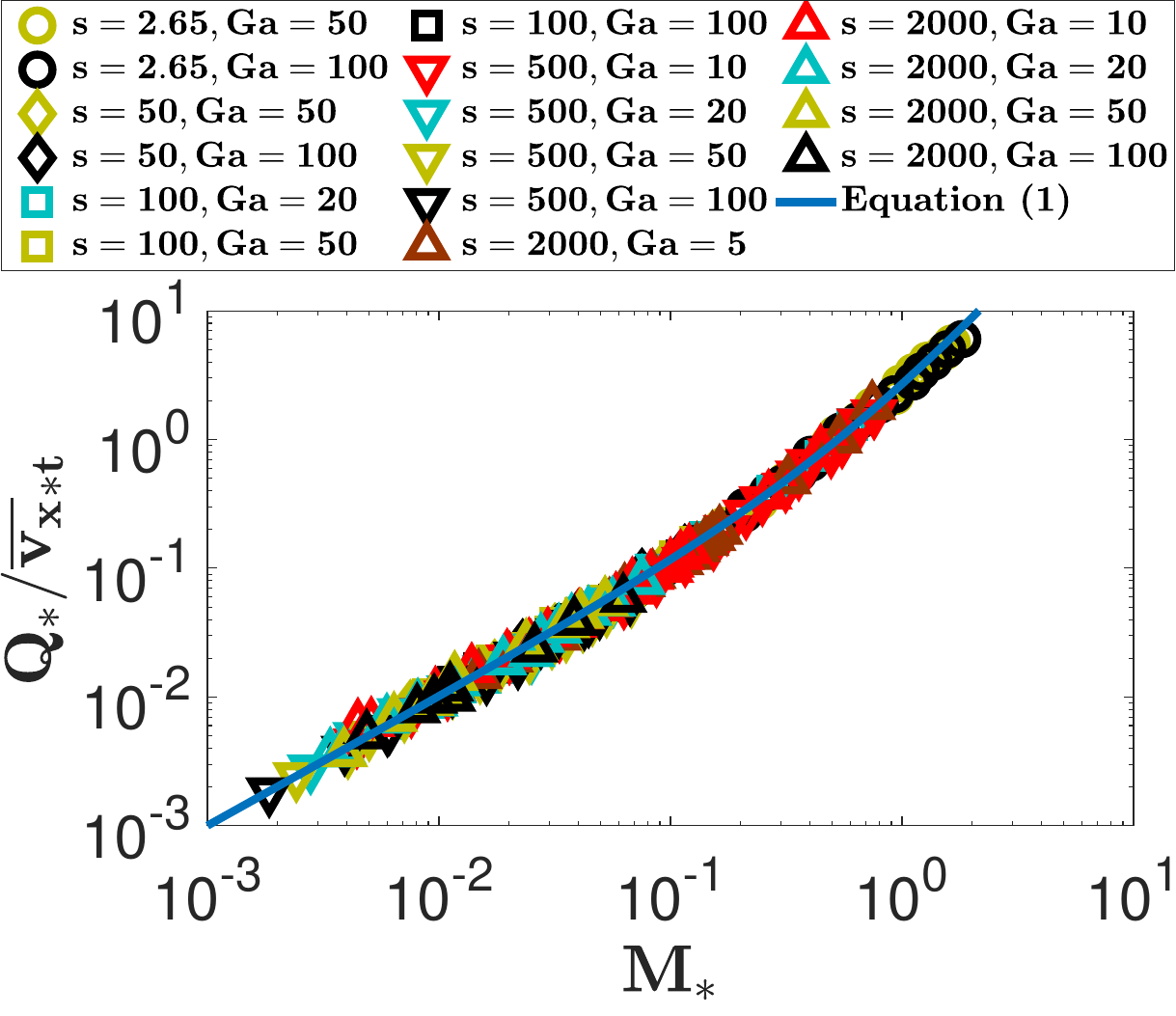}
 \end{center}
 \caption{$Q_\ast/\overline{v_x}_{\ast t}$ versus $M_\ast$. Symbols correspond to data from numerical sediment transport simulations for various combinations of the density ratio $s$, Galileo number $\mathrm{Ga}$, and Shields number $\Theta$, satisfying the conditions $s^{1/2}\mathrm{Ga}\gtrsim80$ for fluvial transport ($s\lesssim 10$) and $s^{1/2}\mathrm{Ga}\gtrsim200$ for aeolian transport ($s\gtrsim 10$). The line corresponds to Eq.~(\ref{Q1}).}
\label{QversusM}
\end{figure}
Using the definitions from the previous paragraph, the main finding of this Letter is the following relationship between $Q_\ast$ and $M_\ast$ (Fig.~\ref{QversusM}):
\begin{equation}
 Q_\ast=M_\ast\overline{v_x}_{\ast t}(1+c_MM_\ast), \label{Q1}
\end{equation}
where $c_M=1.7$ and a quantity's subscript $t$ refers to its value in the limit of no transport: $\Theta\rightarrow\Theta_t(\mathrm{Ga},s)$ (equivalent to $M_\ast\rightarrow0$). The function $\Theta_t(\mathrm{Ga},s)$ defines the transport threshold Shields number for given transport conditions $\mathrm{Ga}$ and $s$~\cite{PahtzDuran18a}. We find that Eq.~(\ref{Q1}) is valid for both fluvial conditions with $s^{1/2}\mathrm{Ga}\gtrsim80$ and aeolian conditions with $s^{1/2}\mathrm{Ga}\gtrsim200$ (Fig.~\ref{QversusM}), encompassing the vast majority of nonsuspended sediment transport conditions that occur in nature, including sand and gravel transport in water and air on Venus, Titan, Earth, Mars, and Pluto. 

Equation~(\ref{Q1}) contains a linear and a quadratic term in $M_\ast$. The linear term dominates when only a few particles are transported ($M_\ast\ll1/c_M$), that is, when the bed can be considered quasistatic and not many binary collisions between transported particles occur. The quadratic term becomes important once $M_\ast\sim1/c_M$, that is, once binary collisions between transported particles become significant. Below, we show that Eq.~(\ref{Q1}) follows naturally from a rigorous link between momentum transport and fluctuation energy dissipation, in which the linear term corresponds to dissipation by fluid drag and particle-bed collisions and the quadratic term to dissipation in binary collisions.

For a single particle of mass $m$ and velocity $\mathbf{v}$ subjected to the force $\mathbf{F}$, Newton's axiom $\mathbf{F}=m\dot{\mathbf{v}}$ dictates $\frac{\d}{\d t}\frac{1}{2}mv_zv_x=F_{(x}v_{z)}$, where the parentheses denote the symmetrization of the indices. This balance relates the horizontal particle momentum to the contact (superscript $c$) and fluid drag (superscript $d$) forces acting on the particle via $\frac{1}{2}mv_x \tilde g = -F^{\tilde g}_{(x}v_{z)}=F^c_{(x}v_{z)}+F^d_{(x}v_{z)}-\frac{\d}{\d t}\frac{1}{2}mv_zv_x$. Summing over all particles per unit bed area and time averaging then yields (steady, homogenous transport conditions)
\begin{equation}
 \frac{1}{2}Q_\ast=D^c_\ast+D^d_\ast, \label{BalanceExz}
\end{equation}
where $(D^c_\ast,D^d_\ast)\equiv(D^c,D^d)/(\rho_p\tilde gd\sqrt{s\tilde gd})$, with $D^c\equiv\frac{1}{\Delta}\langle\sum_iF^{ci}_{(x}v_{z)}\rangle_T$ and $D^d\equiv\frac{1}{\Delta}\langle\sum_iF^{di}_{(x}v_{z)}\rangle_T$ the dissipation rates per unit bed area by particle contact and fluid drag forces, respectively, of $-\frac{1}{\Delta}\langle\sum_i\frac{1}{2}m^iv^i_zv^i_x\rangle_T$~\cite{SuppLiquidBed}, which physically represents a fluctuation energy because $\langle\sum_im^iv^i_z\rangle_T=0$ in the steady state~\cite{Pahtzetal15a}.

\begin{figure*}[!htb]
 \begin{center}
  \includegraphics[width=2.0\columnwidth]{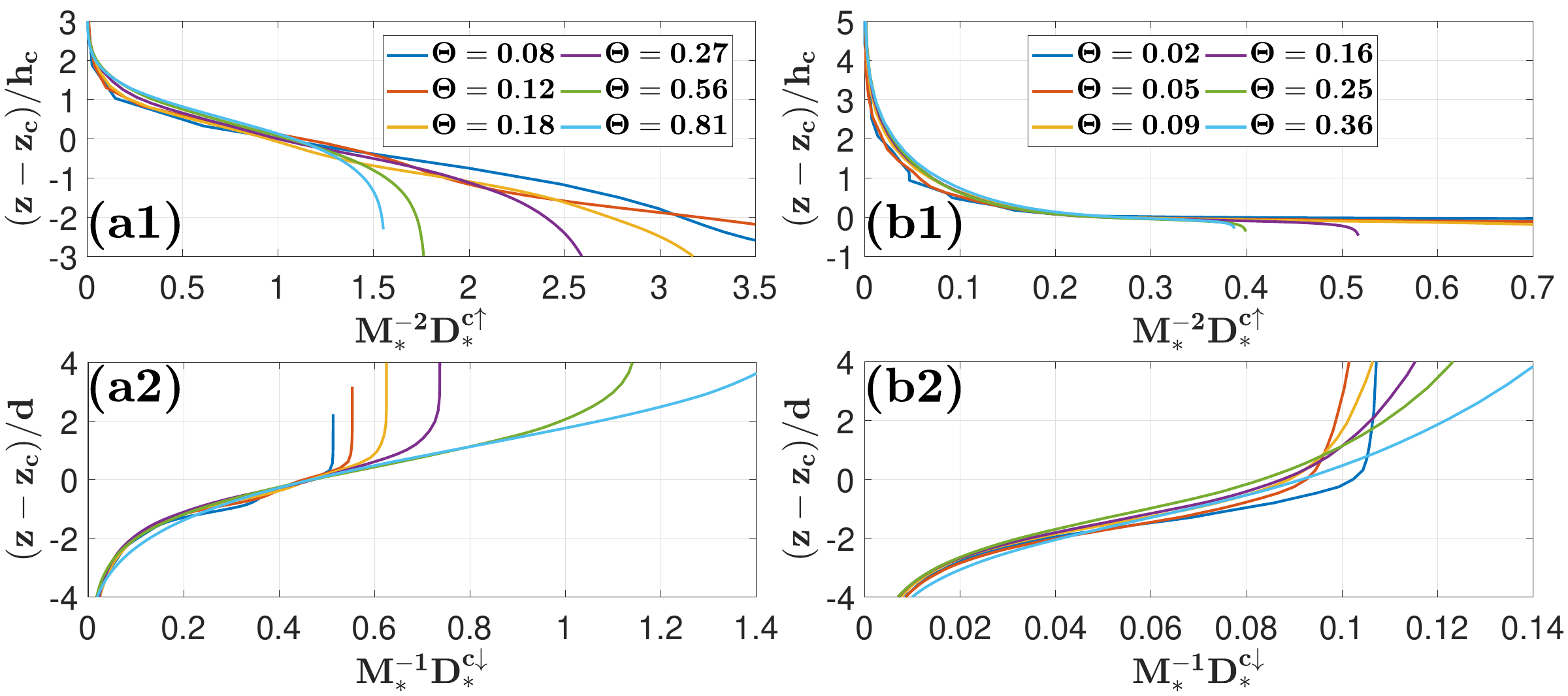}
 \end{center}
 \caption{Vertical profiles of $M_\ast^{-2}D^{c\uparrow}_\ast$ [(a1) and (b1)] and $M_\ast^{-1}D^{c\downarrow}_\ast$ [(a2) and (b2)]. Profiles in (a1) and (b1) are rescaled by the transport layer thickness above $z_c$: $h_c\equiv\int_{z_c}^{\infty}\phi z\d z/\int_{z_c}^{\infty}\phi\d z-z_c$, where $\phi$ is the particle volume fraction. Lines correspond to data from numerical sediment transport simulations for an exemplary fluvial transport condition (density ratio $s=2.65$, Galileo number $\mathrm{Ga}=50$, and various Shields numbers $\Theta$; (a1) and (a2), figure legend in (a1)) and an exemplary aeolian transport condition [$s=2000$, $\mathrm{Ga}=5$, and various $\Theta$; (b1) and (b2), figure legend in (b1)].}
\label{CollisionalDissipation}
\end{figure*}

For any elevation $z$, $D^c$ can be separated into the contact dissipation rate $D^{c\uparrow}(z)\equiv\frac{1}{\Delta}\langle\sum_{z^i\geq z}F^{ci}_{(x}v_{z)}\rangle_T$ of particles moving above $z$ and the contact dissipation rate $D^{c\downarrow}(z)\equiv\frac{1}{\Delta}\langle\sum_{z^i<z}F^{ci}_{(x}v_{z)}\rangle_T$ of particles moving below $z$, that is, $D^c=D^{c\uparrow}(z)+D^{c\downarrow}(z)$. We find that, for any given transport condition, there is an elevation $z_c$ such that $D^{c\uparrow}_\ast(z_c)=a_cM_\ast^2$ and $D^{c\downarrow}_\ast(z_c)=b_cM_\ast$, where $a_c$ and $b_c$ are parameters independent of $M_\ast$ (see Fig.~\ref{CollisionalDissipation} for two exemplary cases). In the region $z\geq z_c$, energy is predominantly dissipated in binary particle collisions, and thus the scaling with $M_\ast^2$ expresses the binary collision probability (analogous to granular kinetic theory~\cite{Kumaran15}). In the region $z<z_c$, the sediment bed is quasistatic and energy dissipation is controlled by the probability of particle-bed collisions, which scales with $M_\ast$. Furthermore, for the relevant transport conditions (legend of Fig.~\ref{QversusM}), we find that the drag dissipation rate $D^d_\ast$ roughly scales with the sediment mass that is responsible for the fluctuation motion (i.e., transported particles), that is, $D^d_\ast=a_dM_\ast$, where $a_d$ is a parameter that is roughly independent of $M_\ast$~\cite{SuppLiquidBed}. Hence, the total dissipation rate ($D^c_\ast+D^d_\ast$) scales as
\begin{equation}
 \frac{1}{2}Q_\ast = D^c_\ast+D^d_\ast=(a_c+a_d)M_\ast+b_cM_\ast^2. \label{BalanceExz2}
\end{equation}
Using the definition $\overline{v_x}_\ast=Q_\ast/M_\ast$ and taking the limit $M_\ast\rightarrow0$ (i.e., $\Theta\rightarrow\Theta_t$), we obtain the closure relation $2(a_c+a_d)=\overline{v_x}_\ast(M_\ast\rightarrow0)\equiv\overline{v_x}_{\ast t}$. Furthermore, the strength of the nonlinear term [$c_M$ in Eq.~(\ref{Q1})] is given by the ratio of the fluctuation energy dissipated in binary collisions and that dissipated by fluid drag and particle-bed collisions: $c_M=b_c/(a_c+a_d)$. Note that, although $b_c$ and $a_c+a_d$ are in general functions of $\mathrm{Ga}$ and $s$, their ratio $c_M$ is approximately constant for relevant conditions (Fig.~\ref{QversusM}).

\begin{figure*}[!htb]
 \begin{center}
  \includegraphics[width=2.0\columnwidth]{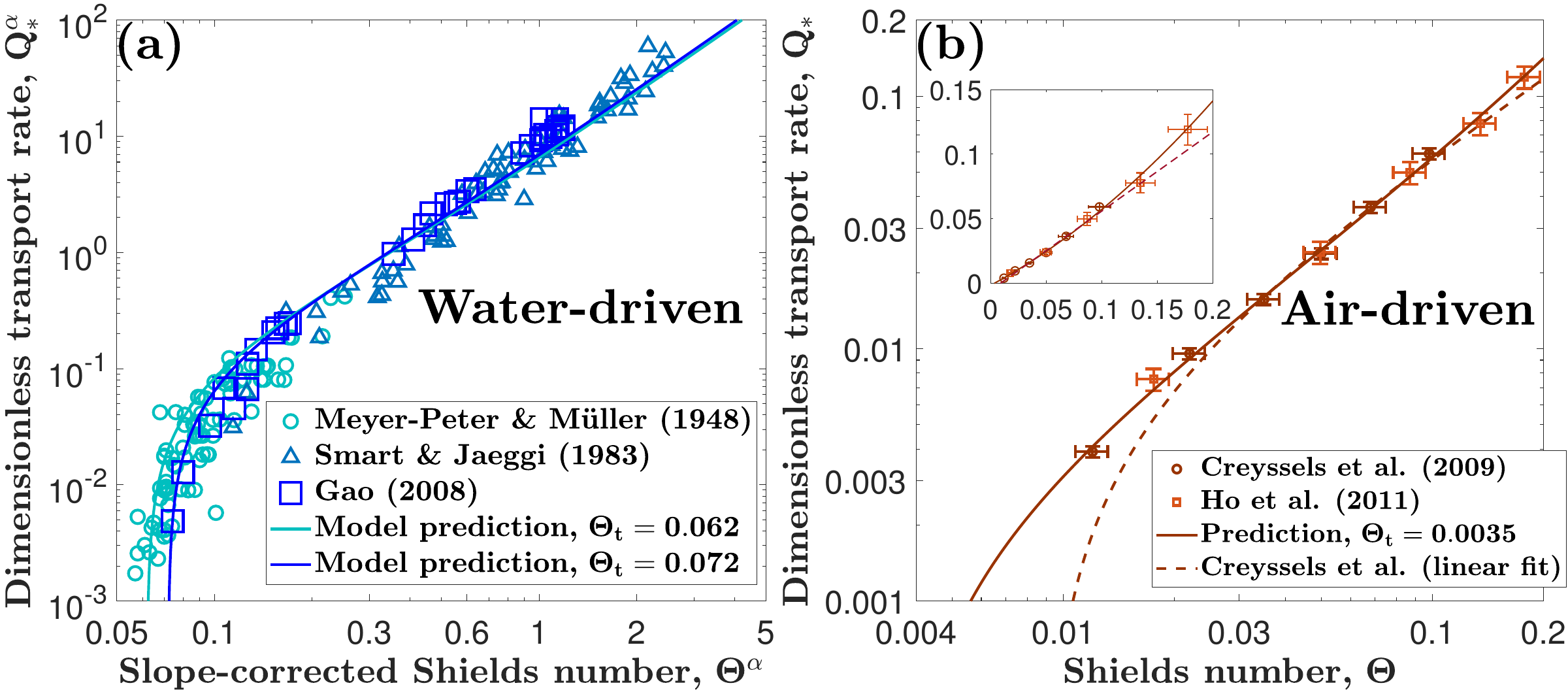}
 \end{center}
 \caption{Test of Eq.~(\ref{Qfinal}) (solid lines) against laboratory measurements of nonsuspended sediment transport driven by (a) water~\cite{MeyerPeterMuller48,SmartJaeggi83,Gao08} and (b) air on Earth~\cite{Creysselsetal09,Hoetal11}. Raw measurements of Ref.~\cite{Gao08} (as reported in Ref.~\cite{Gao03}) and Ref.~\cite{SmartJaeggi83} are corrected for sidewall drag using the method of Ref.~\cite{Williams70} and afterward slope-corrected through Eq.~(\ref{SlopeCorrection}). Measurement data of Ref.~\cite{MeyerPeterMuller48} (mild bed slopes) are as reported in that reference and not further corrected.}
\label{Measurements}
\end{figure*}
\begin{figure}[!htb]
 \begin{center}
  \includegraphics[width=1.0\columnwidth]{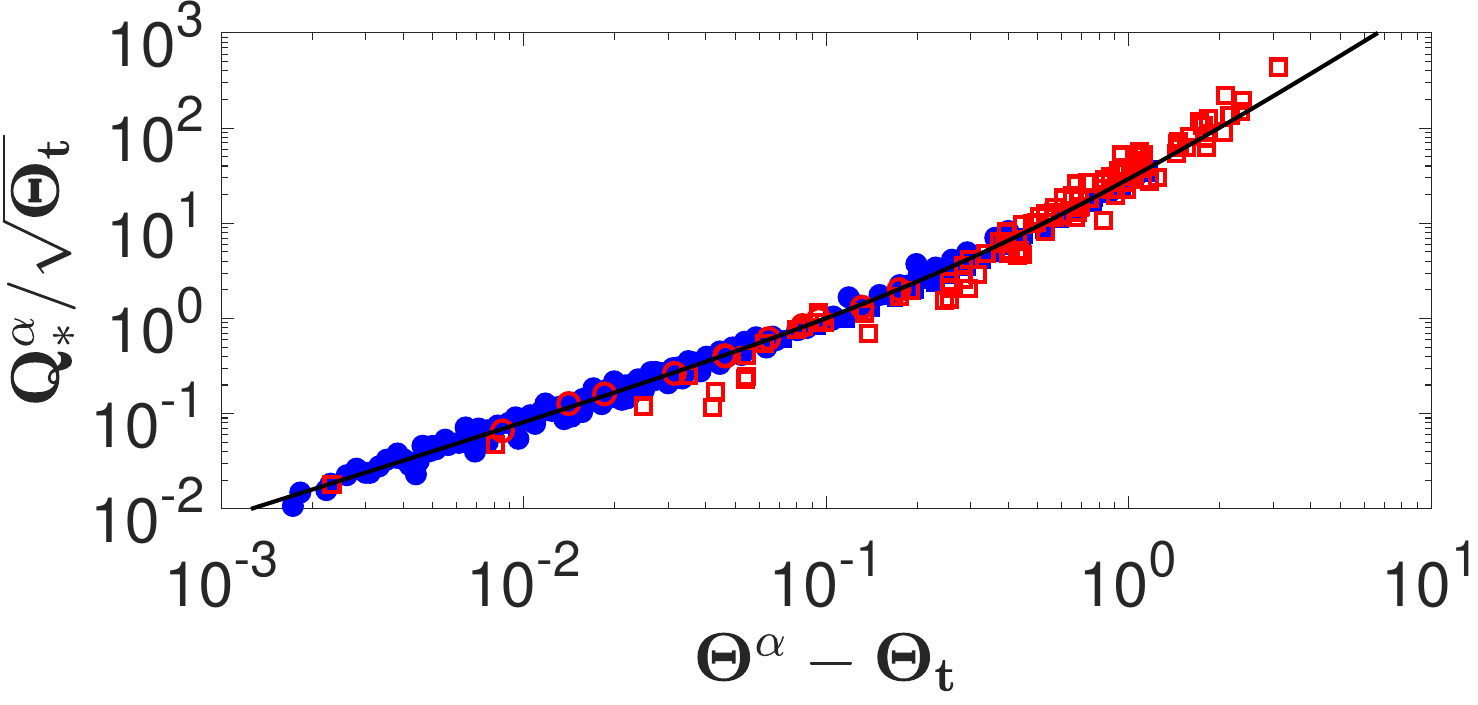}
 \end{center}
 \caption{Test of Eq.~(\ref{Qfinal}) (solid line) against numerical simulations (blue symbols, cf. Fig.~\ref{QversusM}) and laboratory measurements (red symbols, cf. Fig.~\ref{Measurements}) of nonsuspended aeolian (circles) and fluvial (squares) sediment transport. Only those simulation data of Fig.~\ref{QversusM} are shown that obey $s^{1/2}\mathrm{Ga}\gtrsim40$ [validity requirement for Eq.~(\ref{Qfinal})]. The shown fluvial data do not include measurements of Ref.~\cite{MeyerPeterMuller48} for visibility reasons.}
\label{MeasurementsSimulations}
\end{figure}

We further simplify Eq.~(\ref{Q1}) using two insights: (i) steady, homogeneous transport states are those at which $M_\ast$ is so large that the flow (weakened by the particle-flow feedback) is \textit{barely} able to compensate the average energy losses of transported particles rebounding with the bed~\cite{PahtzDuran18b}, and (ii) $\Theta_t$ is the smallest Shields number for which such a state exists~\cite{PahtzDuran18a,Berzietal16}. These insights allow parametrizing $\overline{v_x}_{\ast t}$ and $M_\ast$ in terms of the Shields number $\Theta$ as $\overline{v_x}_{\ast t}=2\kappa^{-1}\sqrt{\Theta_t}$ (valid for $s^{1/4}\mathrm{Ga}\gtrsim40$) and $M_\ast=\mu_b^{-1}(\Theta-\Theta_t)$~\cite{SuppLiquidBed}, where $\kappa=0.4$ is the von K\'arm\'an constant and $\mu_b$ an approximately constant bed friction coefficient that characterizes the geometry of particle-bed rebounds (we choose $\mu_b=0.63$, which yields reasonable overall agreement with the simulation data~\cite{SuppLiquidBed}). Equation~(\ref{Q1}) then becomes
\begin{equation}
 Q_\ast=\frac{2\sqrt{\Theta_t}}{\kappa\mu_b}(\Theta-\Theta_t)\left[1+\frac{c_M}{\mu_b}(\Theta-\Theta_t)\right], \label{Qfinal}
\end{equation}
where we neglect potential effects of particle shape and size distribution on $c_M$ and $\mu_b$.

Equation~(\ref{Qfinal}) exhibits two extreme regimes, a linear scaling $Q_\ast\sim (\Theta-\Theta_t)$ for $(\Theta-\Theta_t)\ll\mu_b/c_M$ (typical for aeolian transport) and a quadratic scaling $Q_\ast\sim\Theta^2$ for $(\Theta-\Theta_t)\gg\mu_b/c_M$ (typical for intense fluvial transport). The latter scaling is consistent with two-phase flow models of intense transport~\cite{PasiniJenkins05,BerziFraccarollo13,Chauchat18,Maurinetal18}. For intermediate values $(\Theta-\Theta_t)\sim\mu_b/c_M$, one can approximate $1+c_M(\Theta-\Theta_t)/\mu_b\approx2\sqrt{c_M(\Theta-\Theta_t)/\mu_b}$, implying $Q_\ast\sim(\Theta-\Theta_t)^{3/2}$, which is one the most widely used scaling laws in hydraulic engineering for the transport of gravel by water~\cite{MeyerPeterMuller48}. Note that Eq.~(\ref{Qfinal}) requires continuous transport conditions to be strictly valid (i.e., $\Theta\gtrsim2\Theta_t$)~\cite{Pahtzetal20a}. In particular, it is expected to underestimate measurements for $\Theta\lesssim\Theta_t$ (for which it predicts $Q_\ast=0$) and overestimate measurements for $\Theta_t\lesssim\Theta\lesssim2\Theta_t$~\cite{Pahtzetal20a}.

In order to compare Eq.~(\ref{Qfinal}), which has been derived for a bed slope angle $\alpha=0$, with slope-driven transport experiments in water (i.e., $\tau=\rho_fgh\sin\alpha$, where $h$ is the clear-water depth), one has to replace $\Theta$ (but not $\Theta_t$) and $Q_\ast^2$ in Eq.~(\ref{Qfinal}) by~\cite{SuppLiquidBed}
\begin{equation}
 (\Theta^\alpha,Q^{\alpha2}_\ast)\equiv(\Theta,Q_\ast^2)\bigg/\left(\cos\alpha-\frac{\sin\alpha}{\mu_b}\frac{s}{s-1}\right). \label{SlopeCorrection}
\end{equation}
When applying this correction and using transport threshold values that are close to (water) or equal to (air) those obtained from a recent threshold model~\cite{PahtzDuran18a}, Eq.~(\ref{Qfinal}) simultaneously reproduces laboratory measurements of the rate of continuous transport in water [Fig.~\ref{Measurements}(a)] and air on Earth [Fig.~\ref{Measurements}(b)]. In particular, the agreement with aeolian transport measurements is substantially better than the old fitted linear model~\cite{Creysselsetal09}. Equation~(\ref{Qfinal}) is also consistent with numerical simulations (Fig.~\ref{MeasurementsSimulations}).

The applicability of Eq.~(\ref{Qfinal}) is limited to transport conditions that exceed critical values of $s^{1/2}\mathrm{Ga}$ and $s^{1/4}\mathrm{Ga}$. The number $s^{1/4}\mathrm{Ga}$ determines whether transported particles tend to move predominantly within the log-layer [$s^{1/4}\mathrm{Ga}\gtrsim40$, as required for Eq.~(\ref{Qfinal})] or within the viscous sublayer of the turbulent boundary layer ($s^{1/4}\mathrm{Ga}\lesssim40$), while $s^{1/2}\mathrm{Ga}$ characterizes the importance of particle inertia relative to viscous drag forcing~\cite{PahtzDuran17,PahtzDuran18a}. When $s^{1/2}\mathrm{Ga}$ is too small [$s^{1/2}\mathrm{Ga}\lesssim80$ for fluvial transport ($s\lesssim10$) and $s^{1/2}\mathrm{Ga}\lesssim200$ for aeolian transport ($s\gtrsim10$)], the scaling of the fluid drag dissipation rate of kinetic particle fluctuation energy changes due to strong viscous drag forcing, causing substantial deviations from Eq.~(\ref{Qfinal}). Likewise, we expect Eq.~(\ref{Qfinal}) to break down when the bed slope angle $\alpha$ comes close to the angle of repose $\alpha_r$, that is, when previous studies suggest that sediment transport properties change relatively abruptly~\cite{Loiseleuxetal05}.

In this Letter, we have shown that the manner in which kinetic particle fluctuation energy is dissipated controls the scaling of the rate $Q$ of continuous nonsuspended sediment transport with the fluid shear stress $\tau$ applied onto the bed. In particular, this scaling becomes nonlinear once fluctuation energy dissipation in binary particle collisions, as opposed to dissipation in particle-bed collisions and via fluid drag, becomes significant. This new physical picture replaces an old, widely-accepted hypothesis about the physical origin of the scaling differences of $Q$. Previously, it was hypothesized that a different predominant mode of bed sediment entrainment is responsible for these differences~\cite{Garcia08,Ancey20,Duranetal11,Koketal12,Andreotti04,Berzietal16}: entrainment caused by the impacts of transported particles onto the bed (``splash''~\cite{Beladjineetal07,ComolaLehning17,Lammeletal17}) in aeolian transport versus entrainment caused by the direct action of fluid forces in fluvial transport. However, a number of recent independent studies revealed that impact entrainment plays a crucial role also in fluvial transport~\cite{Pahtzetal20a,Vowinckeletal16,PahtzDuran17,Clarketal15a,Clarketal17,Heymanetal16,LeeJerolmack18}, making it clear that a better supported hypothesis was needed.

Our physical description has culminated in an expression for $Q$ that unifies transport in water and air streams without fitting to experimental data. In combination with a previous unification of the aeolian and fluvial transport threshold~\cite{PahtzDuran18a}, we are now able to estimate planetary sediment transport and the evolution of planetary sediment surfaces much more reliably than before.

\begin{acknowledgments}
We acknowledge support from grant National Natural Science Foundation of China (Grant No.~11750410687).
\end{acknowledgments}

\renewcommand{\theequation}{S\arabic{equation}}
\renewcommand{\thefigure}{S\arabic{figure}}
\renewcommand{\thetable}{S\arabic{table}}

\section{Supplemental Material}
\subsection{Numerical model details}
The numerical model of Ref.~\cite{Duranetal12} solves the steady, homogeneous Reynolds-averaged Navier-Stokes equations:
\begin{equation}
 \frac{\d\tau_f}{\d z}=f_x,\quad\text{with}\quad\tau_f(\infty)\equiv\tau, \label{NS}
\end{equation}
where $f_x$ is local streamwise force per unit volume that the flow applies on transported particles on average (consisting of the fluid drag, buoyancy, and added mass forces), $z$ the normal-bed coordinate, and $\tau_f$ the Reynolds-averaged fluid shear stress. The latter is calculated using a mixing length model:
\begin{equation}
 \tau_f=\rho_f(1-\phi)\left(\nu+l_m^2\left|\frac{\d u_x}{\d z}\right|\right)\frac{\d u_x}{\d z},
\end{equation}
where $\rho_f$ is the fluid density, $\phi$ is the local particle volume fraction, $u_x$ the Reynolds-averaged streamwise fluid velocity, and $\nu$ the kinematic fluid viscosity. The mixing length $l_m$ is calculated via solving the differential equation
\begin{equation}
 \frac{\d l_m}{\d z}=\kappa\left[1-\exp\left(-\sqrt{\frac{u_xl_m}{R_c\nu}}\right)\right],
\end{equation}
where $\kappa=0.4$ is the von K\'arm\'an constant and $R_c=7$. This expression ensures a smooth hydrodynamic transition from a high $\phi$ within the bed to a low $\phi$ above the bed surface and quantitatively reproduces the mean turbulent flow velocity profile in the absence of transport~\cite{Duranetal12}.

\subsection{Computation of particle properties from the simulation data}
\subsubsection{Local, mass-weighted time average and particle volume fraction}
We compute the local, mass-weighted time average $\langle A\rangle$ of a particle quantity $A$ and particle volume fraction $\phi$ through~\cite{Pahtzetal15a}
\begin{align}
 \langle A\rangle&=\frac{1}{\Delta\phi}\left\langle\sum_nV^nA^n\delta(z-z^n)\right\rangle_T, \label{LocalAverage} \\
 \phi&=\frac{1}{\Delta}\left\langle\sum_nV_p^n\delta(z-z^n)\right\rangle_T, 
\end{align}
respectively, where $\Delta=1181d^2$ is the simulation area, $z^n$ ($V^n=\pi d^{n3}/6$) is the elevation (volume) of particle $n$, $\delta$ the $\delta$ distribution, and $\langle\cdot\rangle_T\equiv\frac{1}{T}\int_0^T\cdot\d t$ denotes the time average over a sufficiently long time $T$. The $\delta$ kernels are coarse grained through spatial averaging over a discretization box of size $1181d\times d\times\Delta z$, where $\Delta z$ varies between $0.05d$ in dense and dilute flow regions ($\phi\gtrsim0.1$) and larger values in rarefied regions. Henceforth, the $\delta$ symbol should thus be interpreted as the associated coarse-graining function.

\subsubsection{Particle stress tensor}
The particle stress tensor $P_{ij}$ is computed through~\cite{Pahtzetal15a}
\begin{equation}
 P_{ij}=\rho_p\phi\langle c_ic_j\rangle+\frac{1}{2\Delta}\left\langle\sum_{mn}F_j^{mn}(x^m_i-x^n_i)K(z,z^m,z^n)\right\rangle_T,
\end{equation}
where $K=\int\limits_0^1\delta\{z-[(z^m-z^n)\tilde s+z^n]\}\d\tilde s$, $\mathbf{c}=\mathbf{v}-\langle\mathbf{v}\rangle$ is the particle fluctuation velocity (with $\mathbf{v}$ the particle velocity), and $\mathbf{F}^{mn}$ the contact force applied by particle $n$ on particle $m$ ($\mathbf{F}^{mm}=0$). We confirmed that these definitions are consistent with the steady, homogeneous local momentum balance $\d P_{zi}/\d z=\rho_p\phi\langle a_i\rangle$~\cite{Pahtzetal15a}, where $\mathbf{a}$ is the particle acceleration due to noncontact forces.

\subsubsection{Fluctuation energy dissipation rate tensors}
The local dissipation rates due to particle contacts and fluid drag, respectively, of $\frac{1}{2}\rho_p\phi\langle c_ic_j\rangle$ are computed through~\cite{Pahtzetal15a}
\begin{subequations}
\begin{align}
 \Gamma^c_{ij}&=-\frac{1}{2\Delta}\left\langle\sum_{mn}F^{mn}_i(v^m_j-v^n_j)\delta(z-z^m)\right\rangle_T, \\
 \Gamma^d_{ij}&=-\rho_p\phi\langle a_ic_j\rangle,
\end{align}
\end{subequations}
respectively.

\subsubsection{Heat flux tensor}
The heat flux tensor, describing the local flux of $\frac{1}{2}\rho_p\phi\langle c_ic_j\rangle$ due to particle contacts and momentum transport between particle contacts, is computed through~\cite{Pahtzetal15a}
\begin{align}
 q_{ijk}&=\frac{\rho_p\phi}{2}\langle c_ic_jc_k\rangle \nonumber \\
 &+\frac{1}{2\Delta}\left\langle\sum_{mn}F^{mn}_jc_k(x^m_i-x^n_i)K(z,z^m,z^n)\right\rangle_T.
\end{align}

\subsection{Numerical model validation}
The numerical model of Ref.~\cite{Duranetal12} reproduces measurements of the transport threshold shear stress (obtained indirectly from Eq.~(\ref{taut})) across aeolian and fluvial nonsuspended sediment transport (Fig.~3 of Ref.~\cite{PahtzDuran18a}). It further reproduces measurements of the vertical profiles of $\phi$ and $\langle v_x\rangle$ in aeolian nonsuspended sediment transport (Fig.~2 of Ref.~\cite{PahtzDuran18a}) and measurements of $\langle v_x\rangle$ in viscous fluvial nonsuspended sediment transport (Fig.~6 of Ref.~\cite{Duranetal14a}). Furthermore, Fig.~(4) of the paper shows that the numerical model also reproduces measurements of the sediment transport rate $Q$.

\subsection{Derivation details}
\subsubsection{Fluctuation energy balance of transported sediment}
In the supplementary material of Ref.~\cite{Pahtzetal15a}, the following general form of the local fluctuation energy balance is derived (Einsteinian summation):
\begin{equation}
 \frac{1}{2}\rho_p\phi\frac{\mathrm{D}\langle c_ic_j\rangle}{\mathrm{D}t}+\frac{\partial q_{k(ij)}}{\partial x_k}=-\frac{P_{k(i}\partial\langle v_{j)}\rangle}{\partial x_k}-\Gamma^d_{(ij)}-\Gamma^c_{(ij)}, \label{FlucEnergyGeneral}
\end{equation}
where $\mathrm{D}/\mathrm{D}t=\partial/\partial t+\langle v_i\rangle\partial/\partial x_i$ is the material derivative and the parentheses denote the symmetrization of the indeces. Equation~(\ref{FlucEnergyGeneral}) describes the local balance of $\frac{1}{2}\rho_p\phi\langle c_ic_j\rangle$ due to production by the mean granular motion, due to dissipation by fluid drag and particle contacts, and due to heat flux. For steady, homogeneous conditions ($\partial_t=\partial_x=\partial_y=0$ and $\langle v_z\rangle=0$~\cite{Pahtzetal15a}), the following two balances can be obtained from Eq.~(\ref{FlucEnergyGeneral}):
\begin{subequations}
\begin{align}
 \frac{\d q_{z(xz)}}{\d z}+\frac{1}{2}P_{zz}\frac{\d\langle v_x\rangle}{\d z}&=-\Gamma^c_{(xz)}-\Gamma^d_{(xz)}, \label{CrossFlucEnergy} \\
 -\frac{\d q^\prime_{zii}}{\d z}-P_{zx}\frac{\d\langle v_x\rangle}{\d z}&=\Gamma^c_{ii}+\Gamma^d_{ii}. \label{FlucEnergy}
\end{align}
\end{subequations}
Integrating these equations over the entire domain ($z\in(\infty,\infty)$) yields
\begin{subequations}
\begin{align}
 -\frac{1}{2}Q\overline{a_z}^q&=-\int\limits_{-\infty}^\infty(\Gamma^c_{(xz)}+\Gamma^d_{(zx)})\d z, \label{IntCrossFlucEnergy} \\
  Q\overline{a_x}^q&=\int\limits_{-\infty}^\infty(\Gamma^c_{ii}+\Gamma^d_{ii})\d z, \label{IntFlucEnergy}
\end{align}
\end{subequations}
where $Q\equiv\int_{-\infty}^\infty\rho_p\phi\langle v_x\rangle\d z=\frac{1}{\Delta}\langle\sum_nm^nv^n_x\rangle_T$ is the definition of the sediment transport rate and $\overline{\cdot}^q\equiv\int_{-\infty}^\infty\langle\cdot\rangle\rho_p\phi\langle v_x\rangle\d z/Q$ a transport flux-weighted height average, and we used the steady, homogeneous local momentum balance $\d P_{zi}/\d z=\rho_p\phi\langle a_i\rangle$~\cite{Pahtzetal15a} and thus
\begin{equation}
 \int\limits_{-\infty}^\infty P_{zi}\frac{\d\langle v_x\rangle}{\d z}\d z=\left[P_{zi}\langle v_x\rangle\right]_{-\infty}^\infty-\int\limits_{-\infty}^\infty\rho_p\phi\langle a_i\rangle\langle v_x\rangle\d z=-Q\overline{a_i}^q.
\end{equation}
We now insert in Eq.~(\ref{IntCrossFlucEnergy}) the definitions
\begin{align}
 D^c&\equiv-\int\limits_{-\infty}^\infty\Gamma^c_{(xz)}\d z, \label{Dc} \\
 D^d&\equiv-\int\limits_{-\infty}^\infty\Gamma^d_{(xz)}\d z+\frac{1}{2}Q(\overline{a_z}^q+\cos\alpha\tilde g), \label{Dd}
\end{align}
where $\tilde g=(1-1/s)g$ is the buoyancy-reduced value of the gravitational constant $g$, with $s\equiv\rho_p/\rho_f$ the particle-fluid-density ratio. This yields a generalization of Eq.~(2) of the paper to arbitrary bed slope angles $\alpha$:
\begin{equation}
 \frac{1}{2}Q_\ast\cos\alpha=D^c_\ast+D^d_\ast,
\end{equation}
where $Q_\ast\equiv Q/(\rho_pd\sqrt{s\tilde gd})$ and $(D^c_\ast,D^d_\ast)\equiv(D^c,D^d)/(\rho_p\tilde gd\sqrt{s\tilde gd})$. However, note that Eq.~(1) of the paper, if generalized to arbitrary $\alpha$, would not change because the same reasoning as in the paper would lead to $2(a_c+a_d)=\cos\alpha\overline{v_x}_{\ast t}$, which means that the additional prefactor $\cos\alpha$ would cancel out.

Equations~(\ref{Dc}) and (\ref{Dd}) reveal that $D^c$ is exactly the dissipation rate due to contacts and $D^d$ approximately (using $-\overline{a_z}^q\simeq\cos\alpha\tilde g$~\cite{PahtzDuran18a}) the dissipation rate due to fluid drag of $-\int_{-\infty}^\infty\rho_p\phi\langle c_zc_z\rangle\d z=-\frac{1}{\Delta}\langle\sum_n\frac{1}{2}m^nv^n_zv^n_x\rangle_T$, as stated in the paper. Furthermore, the ratio $-\overline{a_x}^q/\overline{a_z}^q$ can be interpreted as an effective friction coefficient because of its similarity to the bed friction coefficient $\mu_b\equiv-P_{zx}(z_r)/P_{zz}(z_r)=-\overline{a_x}/\overline{a_z}$, where $\overline{\cdot}\equiv\int_{z_r}^\infty\langle\cdot\rangle\phi\d z/\int_{z_r}^\infty\phi\d z$. This shows that the balance of $-\frac{1}{\Delta}\langle\sum_n\frac{1}{2}m^nv^n_zv^n_x\rangle_T$ stands representative for the balance of the actual fluctuation energy $\frac{1}{\Delta}\langle\sum_n\frac{1}{2}m^n\mathbf{c}^2\rangle_T$ per unit bed area.

\subsubsection{Bed surface elevation}
The fluctuation energy balance plays also a crucial role in the definition of the bed surface elevation $z_r$ (i.e., the effective elevation of energetic particle-bed rebounds). In fact, as particle-bed rebounds are a major reason for the production of $-\frac{1}{2}\phi\langle c_xc_z\rangle$ (Eq.~(\ref{CrossFlucEnergy})), $z_r$ is defined through a maximum of the local production rate~\cite{PahtzDuran18b}:
\begin{equation}
 \max\left(P_{zz}\frac{\d\langle v_x\rangle}{\d z}\right)=\left[P_{zz}\frac{\d\langle v_x\rangle}{\d z}\right](z_r).
\end{equation}

\subsubsection{Momentum balance of transported sediment for arbitrary bed slope angles}
For steady, homogeneous systems, the fluid momentum balance and the horizontal and vertical particle momentum balances read~\cite{Pahtzetal15a}
\begin{subequations}
\begin{align}
 \frac{\d\tau_f}{\d z}&=\rho_p\phi\langle a^d_x\rangle, \label{MomxFluid} \\
 \frac{\d P_{zx}}{\d z}&=\rho_p\phi(\langle a^d_x\rangle+g\sin\alpha), \label{Momx} \\
 \frac{\d P_{zz}}{\d z}&\simeq-\rho_p\phi\tilde g\cos\alpha, \label{Momz}
\end{align}
\end{subequations}
respectively, where $\mathbf{a^d}$ is the drag acceleration (note that $f_x=\rho_p\phi\langle a^d_x\rangle$, see Eq.~(\ref{NS})) and we used that there is no buoyancy force in the flow direction~\cite{Maurinetal18}. In Eq.~(\ref{Momz}), we neglected vertical drag because $|\langle a^d_z\rangle|\ll\tilde g\cos\alpha$~\cite{PahtzDuran18a}. Integrating Eqs.~(\ref{MomxFluid})-(\ref{Momz}) over elevations above the bed surface ($z\in(z_r,\infty)$) and rearranging then yields
\begin{subequations}
\begin{align}
 -P_{zx}(z_r)&=\tau-\tau_f(z_r)+Mg\sin\alpha, \label{IntMomx} \\
 P_{zz}(z_r)&\simeq M\tilde g\cos\alpha, \label{IntMomz}
\end{align}
\end{subequations}
where we used the definition of the transported mass per unit area ($M\equiv\int_{z_r}^\infty\rho_p\phi\d z=\frac{1}{\Delta}\langle\sum_{z^n\geq z_r}m^n\rangle_T$). Our numerical simulations indicate that $\tau-\tau_f(z_r)$ can be well approximated as
\begin{equation}
 \tau-\tau_f(z_r)\simeq\tau-\tau_{t\alpha} \label{taut}
\end{equation}
across sediment transport conditions with $\mathrm{Ga}\sqrt{s}\gtrsim10$~\cite{PahtzDuran18b}, where $\mathrm{Ga}\equiv\sqrt{s\tilde gd^3}/\nu$ is the Galileo number and $\tau_{t\alpha}(\mathrm{Ga},s,\alpha)$ is the slope-modified transport threshold for given values of $\mathrm{Ga}$, $s$, and $\alpha$. Note that the zero-slope transport threshold in the paper is defined as $\tau_t(\mathrm{Ga},s)\equiv\tau_{t\alpha}(\mathrm{Ga},s,0)$ and that we use Eq.~(\ref{taut}) to obtain the transport threshold $\tau_{t\alpha}$ from the simulation data via extrapolation. Equation~(\ref{taut}) is a parametrization of two distinct behaviors of the local fluid shear stress at the surface ($\tau_f(z_r)$): when $\tau$ is close to the transport threshold $\tau_{t\alpha}$, $\tau_f(z_r)$ must also be close to $\tau_{t\alpha}$ because $\tau_f(z_r)\simeq\tau$ when only a few particles are in motion, while for $\tau\gg\tau_{t\alpha}$, the particle shear stress absorbs most of the fluid momentum (i.e., $\tau\gg\tau_f(z_r)$), since net entrainment of bed sediment into the transport layer continuous so long until the near-surface flow is so weak that it can barely sustain a continuous particle motion~\cite{PahtzDuran18b}. Using Eq.~(\ref{taut}) and the definitions $\mu_b\equiv-P_{zx}(z_r)/P_{zz}(z_r)$, $M_\ast=M/(\rho_pd)$, and $\Theta_{(t\alpha)}=\tau_{(t\alpha)}/(\rho_p\tilde gd)$, Eqs.~(\ref{IntMomx}) and (\ref{IntMomz}) can be combined to
\begin{equation}
 M_\ast\simeq\frac{1}{\mu_b\cos\alpha-\frac{s}{s-1}\sin\alpha}(\Theta-\Theta_{t\alpha}). \label{Load}
\end{equation}
For many (but not all) sediment transport conditions, including fluvial conditions, $\Theta_{t\alpha}$ roughly scales as
\begin{equation}
 \Theta_{t\alpha}=\left(\cos\alpha-\frac{\sin\alpha}{\mu_b}\frac{s}{s-1}\right)\Theta_t. \label{ThresholdSlope}
\end{equation}
This scaling is very similar to a slope correction of the transport threshold given in a recent paper~\cite{Maurinetal18}. However, in contrast to the scaling given in Ref.~\cite{Maurinetal18}, $\mu_b$ in Eq.~(\ref{ThresholdSlope}) is not equal to the yield stress ratio $\mu_s$ (using $\mu_s$ instead of $\mu_b$ is inconsistent with the simulation data). Using Eq.~(\ref{ThresholdSlope}) in Eq.~(\ref{Load}), we obtain
\begin{equation}
 M_\ast\simeq\frac{1}{\mu_b}(\Theta^\alpha-\Theta_t), \label{Loadfinal}
\end{equation}
where $\Theta^\alpha\equiv\Theta/(\cos\alpha-\mu_b^{-1}\frac{s}{s-1}\sin\alpha)$.

\subsubsection{Average particle velocity for threshold conditions and arbitrary slope angles}
Following Ref.~\cite{PahtzDuran18a}, the constancy of the bed friction coefficient $\mu_b=-\overline{a_x}/\overline{a_z}\simeq\overline{a_x}/\tilde g$ implies that, for given values of $s$, $\mathrm{Ga}$, and $\alpha$, also $\overline{a^d_x}=\overline{a_x}-g\sin\alpha$ and thus, approximately, the difference between average fluid and particle velocity are constant:
\begin{equation}
 \overline{u_x}-\overline{v_x}=\mathrm{const}. \label{Vr1}
\end{equation}
For near-threshold conditions (i.e., the fluid velocity is not much disturbed by the particle motion), if particles are transported within the log-layer of the turbulent boundary layer, $u_x(z)=\kappa^{-1}\sqrt{s\tilde gd}\sqrt{\Theta}\ln(z/z_o)$ and thus 
\begin{equation}
 \overline{u_x}\simeq\frac{\sqrt{s\tilde gd}\sqrt{\Theta}}{\kappa}\ln\frac{\overline{z}}{z_o},
\end{equation}
where $z_o$ is the surface roughness. Furthermore, when approximating the quasicontinuous particle motion by particles moving in identical periodic trajectories, we can approximate $\overline{z}=c_1v_{\uparrow z}^2/(\tilde g\cos\alpha)$ and $\overline{v_x}=c_2v_{\uparrow z}$, where $v_{\uparrow z}$ is the vertical lift-off velocity of the particles and $c_1$ and $c_2$ are proportionality constants~\cite{Berzietal16}. Hence, Eq.~(\ref{Vr1}) becomes
\begin{equation}
 \frac{\sqrt{s\tilde gd}\sqrt{\Theta}}{\kappa}\ln\frac{c_1v_{\uparrow z}^2}{\tilde gz_o\cos\alpha}-c_2v_{\uparrow z}=\mathrm{const}. \label{Vr2}
\end{equation}
Now, the transport threshold $\Theta_{t\alpha}$ corresponds to the smallest value of $\Theta$ as a function of $v_{\uparrow z}$ that allows for continuous particle trajectories (i.e., the smallest $\Theta$ for which a solution of Eq.~(\ref{Vr2}) exists), which implies $[\d\Theta/\d v_{\uparrow z}]_t=0$~\cite{PahtzDuran18a}. Hence, evaluating the derivative of Eq.~(\ref{Vr2}) with respect to $v_{\uparrow z}$ at threshold conditions yields
\begin{equation}
 \overline{v_x}_{\ast t}\equiv\frac{\overline{v_x}_t}{\sqrt{s\tilde gd}}=\frac{c_2v_{\uparrow z}}{\sqrt{s\tilde gd}}=\frac{2}{\kappa}\sqrt{\Theta_{t\alpha}}. \label{vxt}
\end{equation}
Inserting Eqs.~(\ref{Loadfinal}) and (\ref{vxt}) into Eq.~(1) of the paper, using $Q_\ast\equiv Q/(\rho_pd\sqrt{s\tilde gd})=M_\ast\overline{v_x}_\ast$, then yields
\begin{equation}
 Q_\ast=\frac{2\sqrt{\Theta_{t\alpha}}}{\kappa\mu_b}(\Theta^\alpha-\Theta_t)\left[1+\frac{c_M}{\mu_b}(\Theta^\alpha-\Theta_t)\right].
\end{equation}
Finally, using Eq.~(\ref{ThresholdSlope}) (valid for fluvial transport conditions), we obtain
\begin{equation}
 Q^\alpha_\ast=\frac{2\sqrt{\Theta_t}}{\kappa\mu_b}(\Theta^\alpha-\Theta_t)\left[1+\frac{c_M}{\mu_b}(\Theta^\alpha-\Theta_t)\right],
\end{equation}
where
\begin{equation}
 Q^\alpha_\ast\equiv\frac{Q_\ast}{\sqrt{\Theta_{t\alpha}/\Theta_t}}=\frac{Q_\ast}{\sqrt{\cos\alpha-\frac{\sin\alpha}{\mu_b}\frac{s}{s-1}}}.
\end{equation}
This explains the slope correction (Eq.~(5) of the paper).

\begin{figure*}[!htb]
 \begin{center}
  \includegraphics[width=2.0\columnwidth]{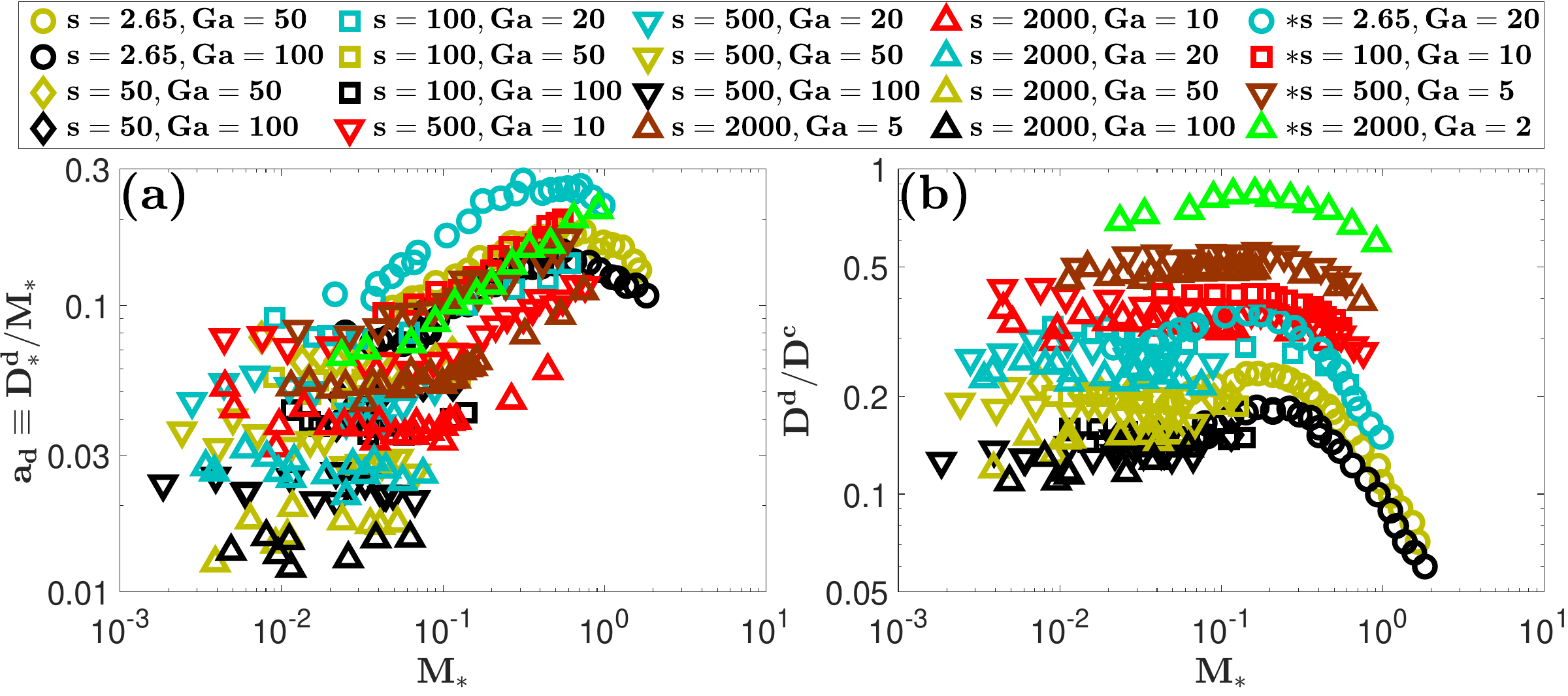}
 \end{center}
 \caption{(a) Scaling parameter $a_d\equiv D^d_\ast/M_\ast$ versus dimensionless sediment transport load $M_\ast$. (b) Ratio between the fluctuation energy dissipation rate due to fluid drag forces and the dissipation rate due to contact forces ($D^d/D^c$) versus $M_\ast$. Symbols correspond to data from sediment transport simulations for various combinations of the density ratio $s$, Galileo number $\mathrm{Ga}$, and Shields number $\Theta$. The asterisks in the figure legend indicate fluvial conditions ($s\lesssim10$) with $s^{1/2}\mathrm{Ga}\lesssim80$ or aeolian conditions ($s\gtrsim10$) with $s^{1/2}\mathrm{Ga}\lesssim200$.}
\label{DragDissipation}
\end{figure*}
\begin{figure*}[!htb]
 \begin{center}
  \includegraphics[width=2.0\columnwidth]{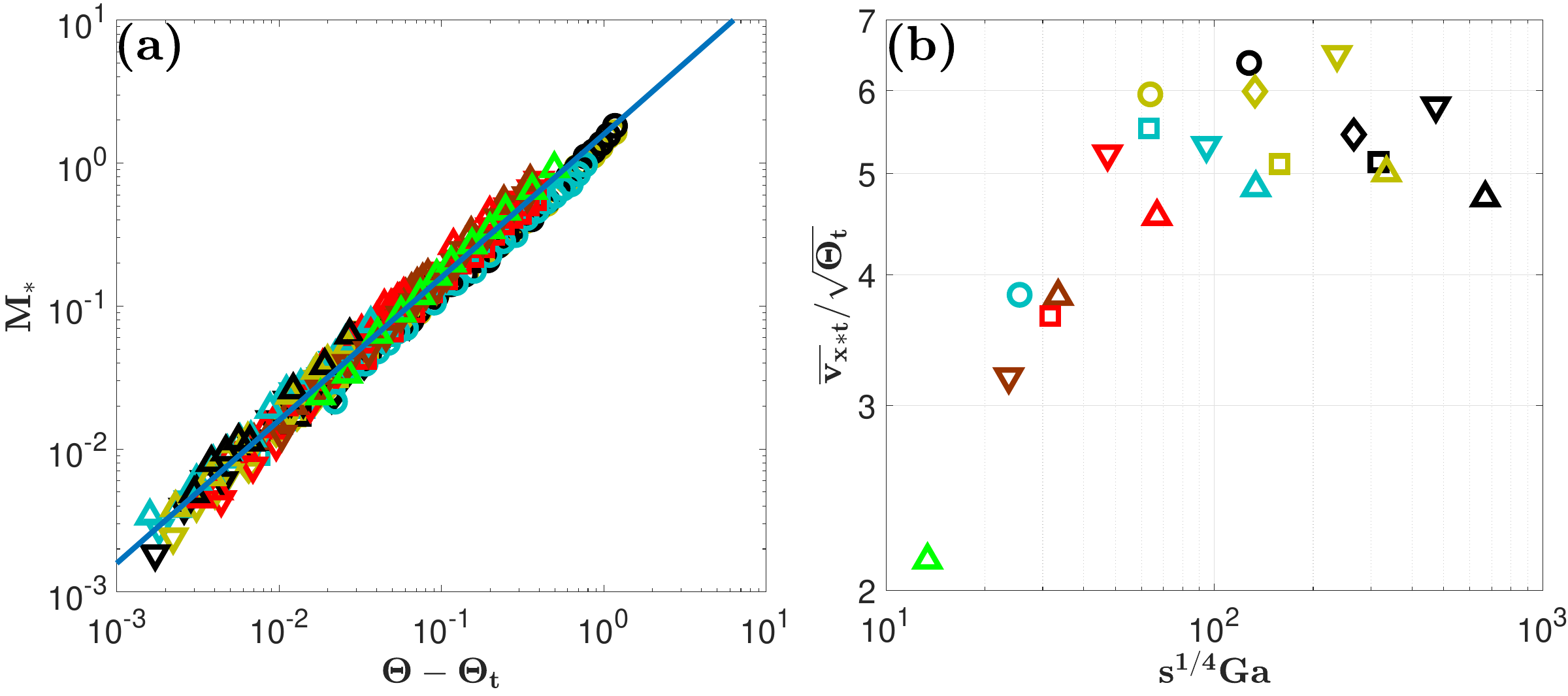}
 \end{center}
 \caption{(a) Dimensionless sediment transport load $M_\ast$ versus Shields number in excess of the threshold ($\Theta-\Theta_t$). (b) Rescaled average horizontal particle velocity ($\overline{v_x}_{\ast t}/\sqrt{\Theta_t}$) versus $s^{1/4}\mathrm{Ga}$. Symbols correspond to data from sediment transport simulations for nonsloped beds ($\alpha=0$) and various combinations of $s$, $\mathrm{Ga}$, and $\Theta=\Theta^\alpha|_{\alpha=0}$. The asterisks in the figure legend indicate fluvial conditions ($s\lesssim10$) with $s^{1/2}\mathrm{Ga}\lesssim80$ or aeolian conditions ($s\gtrsim10$) with $s^{1/2}\mathrm{Ga}\lesssim200$. The line in (a) corresponds to Eq.~(\ref{Loadfinal}) with $\mu_b=0.63$.}
\label{ValidationSimulations}
\end{figure*}
\begin{figure*}[!htb]
 \begin{center}
  \includegraphics[width=2.0\columnwidth]{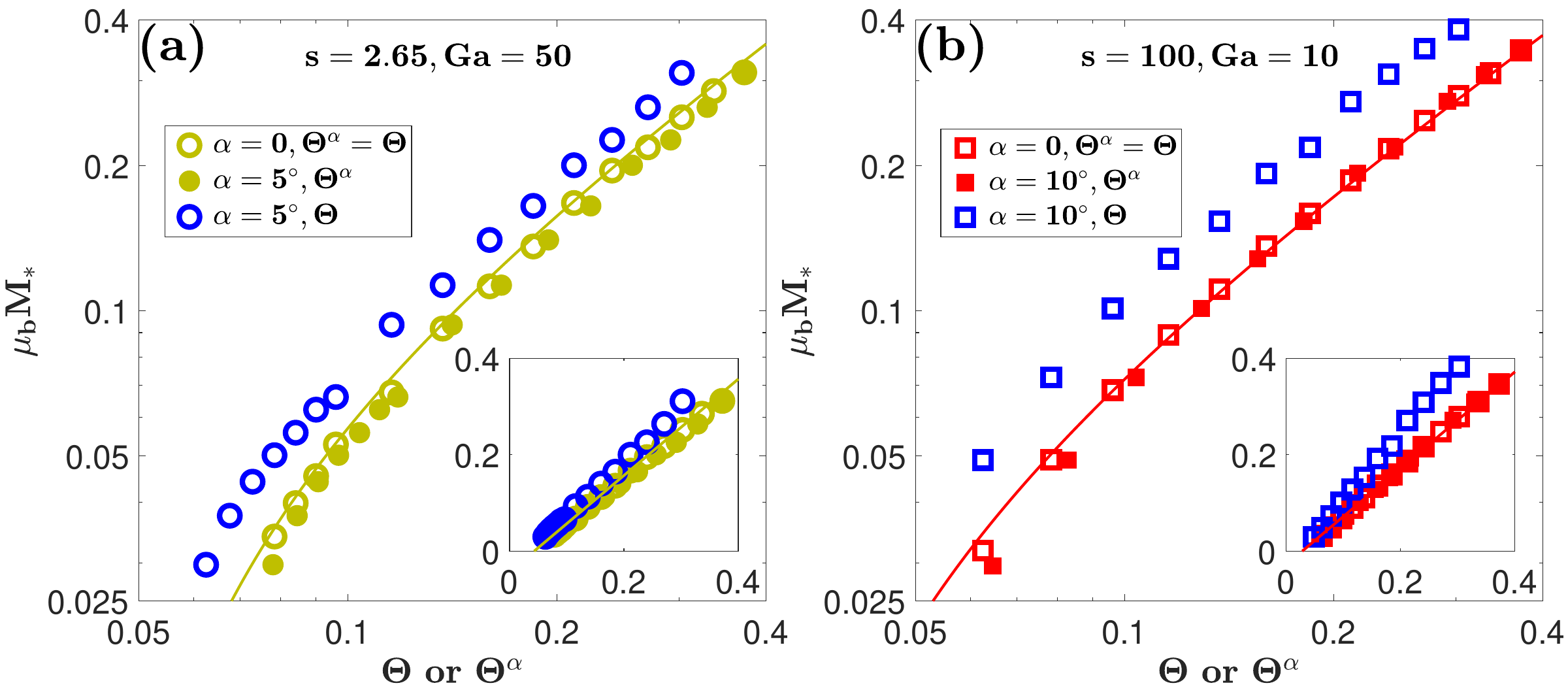}
 \end{center}
 \caption{Rescaled dimensionless transport load $\mu_bM_\ast$ versus slope-corrected Shields number $\Theta^\alpha$ or versus Shields number $\Theta$ (only for blue symbols). Symbols correspond to data from sediment transport simulations and (a) $s=2.65$, $\mathrm{Ga}=50$, $\alpha=[0,5^\circ]$, and various $\Theta$, or (b) $s=100$, $\mathrm{Ga}=10$, $\alpha=[0,10^\circ]$, and various $\Theta$. Lines correspond to Eq.~(\ref{Loadfinal}).}
\label{SlopeSimulations}
\end{figure*}
\subsection{Scaling of the fluid drag dissipation rate}
Figure~\ref{DragDissipation}(a) shows that, for fluvial ($s\lesssim10$) transport conditions with $s^{1/2}\mathrm{Ga}\gtrsim80$ and aeolian ($s\gtrsim10$) conditions with $s^{1/2}\mathrm{Ga}\gtrsim200$, the drag dissipation scales roughly as $D^d_\ast\propto M_\ast$ (i.e., $a_d\equiv D^d_\ast/M_\ast\approx\mathrm{const}$). Deviations from this scaling are typically found for large $M_\ast$, for which, however, the contribution of $D^d$ relative to the contact dissipation rate $D^c$ becomes relatively small (Fig.~\ref{DragDissipation}(b)). Only for fluvial conditions with $s^{1/2}\mathrm{Ga}\lesssim80$ and aeolian conditions with $s^{1/2}\mathrm{Ga}\lesssim200$ (those indicated by asterisks) do these deviation begin to matter (i.e., they cause deviations from Eq.~(1) in the paper).

\subsection{Validation of Eqs.~(\ref{Loadfinal}) and (\ref{vxt})}
Figure~\ref{ValidationSimulations}(a) shows that Eq.~(\ref{Loadfinal}) with $\mu_b=0.63$ is approximately valid across all simulated conditions with $\alpha=0$. Figure~\ref{ValidationSimulations}(b) shows that Eq.~(\ref{vxt}) is approximately valid once $s^{1/4}\mathrm{Ga}\gtrsim40$, which is consistent with the transition to a transport regime in which most particles are transported within the log-layer of the turbulent boundary layer, which is the requirement for the validity of Eq.~(\ref{vxt})~\cite{PahtzDuran18a}. Figure~\ref{SlopeSimulations} shows for two exemplary cases, one of which corresponds to fluvial sediment transport, that the data for different bed slope angles $\alpha$ obey Eq.~(\ref{Loadfinal}). Figure~\ref{SlopeSimulations} also shows that the effect of $\alpha$, in general, cannot be neglected (i.e., the approximation $\Theta^\alpha\approx\Theta$ does not work well).

%

\end{document}